\documentclass[letterpaper,twocolumn,10pt]{article}
\usepackage{usenix,epsfig}
\usepackage[hyphens]{url}
\usepackage{hyperref}
\usepackage[utf8]{inputenc}
\begin{document}

\date{}

\title{\Large \bf Fingerprintability of WebRTC}

\author{
{\rm David Fifield\thanks{Authors are listed in alphabetical order.}}\\
University of California, Berkeley
\and
{\rm Mia Gil Epner}\\
University of California, Berkeley
}

\maketitle

\thispagestyle{empty}

\subsection*{Abstract}

We examine WebRTC's suitability as a means of
Internet censorship circumvention.
WebRTC is a framework and suite of protocols
for peer-to-peer communication between web browsers.
We analyze the implementation differences
in instantiations of WebRTC that make it possible to
``fingerprint'' implementations---potentially distinguishing circumvention-related uses
from ordinary ones.
This question is relevant to Snowflake,
an upcoming circumvention system
that uses WebRTC to turn web browsers
into temporary peer-to-peer proxies.
We conduct a manual analysis of WebRTC-using applications
in order to map the space of distinguishing implementation features.
We run a fingerprinting script on a day's worth of network
traffic in order to quantify WebRTC's prevalence and diversity.
Throughout, we find pitfalls that indicate
that resisting fingerprinting in WebRTC is likely to be non-trivial.

\section{Background}

The job of an Internet censor is essentially that of traffic classification.
The censor observes traffic and decides,
perhaps on a per-packet or per-stream basis,
whether to block or allow it.
The censor incurs a cost whenever it classifies incorrectly,
whether by overblocking
(blocking what what should be allowed,
thereby diminishing the utility of the Internet),
or by underblocking
(allowing what should be blocked, thus failing in the task of censorship).
Circumvention attempts to increase the difficulty of the classification problem,
by making forbidden traffic look like allowed traffic
and causing the censor to misclassify more often.
This task
requires that the forbidden traffic resemble the allowed traffic
in every aspect that is easily observable to the censor---otherwise
the censor can easily separate the wheat from the chaff.

Even a naive censor will block direct access
to forbidden resources, for example web sites.
Therefore circumvention systems typically employ some kind of proxy
that provides indirect access to blocked resources.
The challenge for the censor, then, is to discover and block
communication with proxies.
Perhaps the biggest challenge in disguising proxy access
is camouflaging addressing information, such as the destination IP addresses.
The IP addresses of proxies must not be easily distinguishable
as belonging to circumvention,
or else the censor's classification task is very easy:
it can decide what to block or allow based on nothing more than the destination.
There are several systems that aim to solve this IP-blocking problem,
The one we focus on is Snowflake~\cite{snowflake}, now under development,
which takes the approach of running proxies inside of web browsers.
The browsers serve as a source of cheap and unpredictable IP addresses.

Suppose that Snowflake adequately solves the problem of IP blocking.
There remains another challenge,
network protocol classification.
which is classification by the network protocol used to
communicate with proxies.
Snowflake is based on WebRTC~\cite{ietf-rtcweb-overview-15},
a peer-to-peer framework built into web browsers.
The question we address is whether Snowflake's use of WebRTC protocols
stands out from other applications',
and whether there is enough use of WebRTC in the wild
that a censor cannot easily just block it wholesale.

Snowflake has two components that use WebRTC:
the proxy, implemented in JavaScript for a browser;
and the client, which uses a headless standalone implementation of WebRTC.
Because of the requirements of the circumvention design,
when establishing a WebRTC connection,
the proxy is the initiator and the client is the receiver;
i.e., the proxy plays the role of a WebRTC client
and the client plays the role of a server.

\subsection{WebRTC}

WebRTC is a rather complicated set of protocols and an API for
building communications applications.
It was originally designed for applications such as voice and video
chat---the ``RTC'' stands for real-time communications.
In addition to the transport of media streams,
WebRTC offers TCP-like reliable and UDP-like unreliable data channels.
WebRTC has only recently become commonly available
and well supported in web browsers.
What makes WebRTC interesting for circumvention
is its built-in support for traversal of
network address translation (NAT).
A lack of NAT traversal was a major impediment
to the adoption of flash proxy~\cite{Fifield2012a},
an earlier, TCP-based system that also used in-browser proxies.

WebRTC incorporates a bundle of related protocols.
It uses Interactive Connectivity Establishment (ICE)~\cite{rfc5245}
for NAT traversal.
ICE itself uses the NAT-traversal technologies
STUN~\cite{rfc5389} and TURN~\cite{rfc5766}.
Data channels are implemented as
a transport layer atop Datagram TLS (DTLS)~\cite{rfc6347}.
Media streams are carried over the
Secure Real-time Transport Protocol (SRTP)\cite{rfc3711},
however usually using a DTLS key exchange in a process known as
DTLS-SRTP~\cite{rfc5764}.
Some WebRTC applications that use SRTP make use of
an older type of key exchange called SDES~\cite{rfc4568}---in
this case no DTLS handshake occurs.

Our work is a preliminary step towards anticipating how successful
WebRTC-based circumvention is likely to be.
We investigate these questions:
\begin{enumerate}
\item How much WebRTC exists in the wild?
\item What kind of diversity is there in naturally occurring WebRTC fingerprints?
\item Is it feasible to mimic the fingerprint of an application that is important enough to resist blocking?
\end{enumerate}

\subsection{Threat model}

In our threat model,
the censor controls a perimeter around a censored user.
All a user's communication is mediated by the censor,
who decides what to block or allow according to its own policy and capabilities.
The censor, however, does not control the user's personal computer.
The user's goal is to reach some blocked network resource
outside the censor's control.
The censor is free to block, modify, replay, or inject
any traffic according to its own wishes.
We assume, however, that the censor seeks to avoid
overblocking; the potential for circumvention
increases the more this assumption holds.

\section{Candidate features}

Here we identify a number of WebRTC protocol features
that we expect to be useful for fingerprinting.

\begin{description}
\item[STUN and TURN]
The STUN and TURN NAT-traversal protocol are valuable in several ways.
Messages contain a list of attributes whose order and contents are left up to the implementer,
including a SOFTWARE attribute that explicitly identifies
the implementation, like the User-Agent header in HTTP.
Not only client traffic but also server traffic is distinguishable.
We can evaluate the type of server that the application connects to;
also, the selection of what STUN servers to use is a choice made by the client application.
The type of packets being sent can be used as a fingerprint.
The majority of applications send only Binding requests, successful Binding responses, Allocate requests, and successful Allocate responses.
The minority send, in addition, CreatePermission requests and responses, and send indication packets.
Some applications force the use of UDP relaying using TURN, against the guidelines of WebRTC.
\item[DTLS]
The DTLS layers has several features that contribute to fingerprinting,
mostly inherited from TLS.
These include the DTLS version (DTLSv1.0 and DTLSv1.2 are the possibilities),
the ordered lists of cipher suites and extensions offered by the client,
the cipher suite chosen by the servers, and the server's extensions.
The certificate offered by the server has interesting details too,
including the ``common name'' field and the period of validity.
\item[Media vs.~data transport]
Snowflake, at its current stage of development,
always uses reliable WebRTC data channels,
meaning that the on-the-wire protocol seen by the censor is DTLS.
Other WebRTC-based applications use media channels,
which use DTLS-SRTP or SRTP with SDES.
Though all these protocol are encrypted,
it is easy to distinguish one from another.
\end{description}

\section{Prior work} 

Nick Mathewson wrote in 2012 on the difficulty of
disguising Tor's TLS connections~\cite{tor-tlshistory}:
\begin{quote}
``At this point, we hadn't actually learned very much about TLS internals:
we were treating TLS as an idealized black-block encrypted transport. Obviously, this was a mistake on our part.''
\end{quote}
This early lack of understanding was the cause
of some regrettable design decisions:
Tor's use of TLS is more complicated than necessary
because of incrementally added fingerprinting mitigations.
Our research aims to pre-empt these difficulties
when building with WebRTC,
by understanding the issues thoroughly at the start.

DTLS is an adaptation of TLS for datagram transports,
and therefore inherits the fingerprintability of TLS.
Majkowski~\cite{p0f-ssl} built a TLS client fingerprinting plugin
for p0f, the passive OS fingerprinting tool.
It used as features the TLS version, client ciphersuites and extensions,
and other implementation quirks.
Fifield et~al.~\cite{Fifield2015a} emphasized the importance
of matching a browser's fingerprint when using HTTPS
for circumvention.

Houmansadr et~al. in their influential
``parrot is dead'' paper~\cite{Houmansadr2013b}
argue that superficial protocol imitation
is fundamentally flawed,
due to the great many details
one must get exactly right in order to remain indistinguishable.
Subtle protocol details, such as endpoints'
behavior in the face of errors,
are enough to unmask naive protocol mimicry.
As WebRTC is a large framework consisting of
complicated protocols, imitation by mimicry
should be especially infeasible.

uProxy~\cite{uproxy} is a circumvention system that can,
among other things,
route censored users' traffic through the browsers
of their friends using data channels and WebRTC.
uProxy can additionally obfuscate the DTLS layer
using transformation programs that hide the fact that DTLS is in use~\cite{uproxy-obfuscators}.
This is possible for uProxy, and not for Snowflake,
because uProxy is a browser extension
that has extra capabilities compared to an ordinary web application.

The website \href{https://webrtchacks.com/}{webrtchacks.com}
has done reverse-engineering and discussion
of WebRTC applications,
which we found useful in our own investigation.

\section{Manual fingerprint analysis}
We began by analyzing several WebRTC implementations in web applications. Using Wireshark to capture the traffic, we attempted to discover notable features or idiosyncrasies in these implementations. We studied the DTLS connections, as well the STUN/TURN packets and DNS lookups of STUN/TURN servers, by manually analyzing the traces.

We analyzed traces from browser-to-browser Facebook Messenger, Google Hangouts, OpenTokRTC, Sharefest, and Snowflake. We chose Facebook Messenger and Google Hangouts because they are popular applications that were discovered~\cite{webrtchacks-facebook,webrtchacks-hangouts} to be using WebRTC. We chose OpenTokRTC because of its advertised usage of WebRTC. We chose to study Sharefest as it is a data-only connection, rather than the voice and video services of the prior applications.
\subsection{Google Hangouts}
Google Hangouts (\url{https://hangouts.google.com/}) video chat begins with STUN binding requests made to the Google STUN server. This is followed by Binding success responses. Through the \begin{NoHyper}\url{chrome://webrtc-internals}\end{NoHyper} display in Chrome, we see ``DtlsSrtpKeyAgreement:false'', meaning that the key is exchanged through SDES rather than through DTLS. DTLS is not present in this implementation. 
\subsection{Facebook Messenger}
Facebook Messenger (\url{https://www.messenger.com/}) uses WebRTC with DTLS for browser-to-browser communication but uses WebRTC with SDES for any communication involving a mobile device. Messenger begins with Binding requests sent to both a Facebook STUN server and a Facebook TURN server, but then only sends Allocate requests and CreatePermission requests to the TURN server, indicating that Facebook has forced TURN usage. Several additional send indication TURN packets are sent, some over TCP, others over UDP. These are used to forward data to a peer through the TURN server.

Next, the DTLS connection begins. The DTLS client hello contained several potentially fingerprintable attributes: DTLSv1.0 was used, nine cipher suites were offered, there was a null compression method, the use\_srtp extension was present, and there were two elliptic curves offered. The server hello responds with the cipher suite TLS\_ECDHE\_RSA\_WITH\_AES\_256\_CBC\_SHA. The server's certificate includes the common name ``WebRTC''. The certificate has a validity period of 30 days. The connection continues with an SRTP-based connection.
\subsection{OpenTokRTC}
OpenTokRTC (\url{https://opentokrtc.com/}) is a WebRTC-based chat demo.
WebRTC begins in OpenTok with STUN binding requests and successes to TURN Tokbox servers. Several Allocate requests follow, many of which error. The error response code is consistently 401, meaning unauthorized. The response packets include two notable attributes. First, they carry a REALM attribute with the contents tokbox.com. Second, their SOFTWARE attribute identifies the server as ``Citrix-3.2.5.1 `Marshal West'''. This is defined as an identifier for a free TURN server~\cite{rfc5766-turn-server}. When the Allocate requests succeed, they provide a username.

The DTLS connection begins with DTLSv1.0. 73 cipher suites are offered by the client, many of them outdated and attackable. We do not expect this to appear on other WebRTC DTLS connections. The hello also includes a null compression method, the use\_srtp extension, and a heartbeat extension. The server hello chose cipher suite TLS\_ECDHE\_RSA\_WITH\_AES\_256\_CBC\_SHA. The server key was exchanged using elliptic key curve secp256r1. The certificate includes the common name ``WebRTC''. The certificate also had a validity period of 30 days. Following the establishment of the DTLS connection, the video chat continued over SRTP. 

\subsection{Sharefest}
Sharefest (\url{https://sharefest.me/}) is a file sharing service that uses WebRTC to transmit over a data channel. Our trace of Sharefest began with a STUN connection to one of Google's STUN server. Only STUN binding requests and binding successes were sent. 

The Sharefest DTLS connection began with two client hellos. The two hellos were identical, except for different IP header identification values and different sequence numbers. The client used DTLSv1.0, offered nine different cipher suites, a null compression method, the use\_srtp extension, and two elliptic curves. The server hello chose TLS\_ECDHE\_RSA\_WITH\_AES\_256\_CBC\_SHA, and included the use\_srtp extension. The server certificate has the common name ``WebRTC'' and a validity period of exactly 30 days. Additionally, the named elliptic curve was secp256r1. The DTLS connection transmitted the entirety of the data, which differed from the previous services.
\subsection{Snowflake}
Snowflake begins its WebRTC connection with STUN Binding requests sent to Google's STUN server. The STUN Binding requests and success responses continue. A DTLS connection begins with a client hello with DTLSv1.0 and server hello version of DTLSv1.2. This is the first connection we have seen willing to support DTLSv1.2. The client hello offers 17 cipher suites. The client hello also offers a null compression method, a signature algorithms extension, the use\_srtp extension, and the renegotiation info extension.

The DTLS server hello chooses DTLSv1.2. Snowflake is the only one of the applications we analyzed to use version 1.2. The server hello chose cipher suite TLS\_ECDHE\_RSA\_WITH\_AES\_128\_GCM\_SHA256, which is distinct from all of the other WebRTC applications. Also included was the use\_srtp extension. The server certificate includes the common name ``WebRTC'' with a validity period of 30 days. 
\subsection{Observations}
Manual analysis gave us a list of factors that could influence a DTLS fingerprint. These include the list of client extensions, the cipher suites and elliptic curves, the certificate validity and common name. We discovered that Facebook Messenger and Google Hangouts did not use WebRTC for text-based chats, only for video chats. We also determined that Hangouts did not use DTLS to exchange keys. 
 

\section{DTLS fingerprinting in a large traffic trace}

We wrote a DTLS fingerprinting script for Bro~\cite{bro}.
For every DTLS handshake, the script captures the timestamp; a unique ID;
the DTLS version;
the client's lists of cipher suites, extensions, and elliptic curves;
the server's chosen cipher suite, elliptic curve, compression method, and list of extensions;
and the interval of validity of the server certificate.
We combine these features into a fingerprint
consisting of a long text string.
The script records a log line whenever
a DTLS connection is successfully established,
or when there is a TLS alert message terminating the handshake.

The script captures only one part of the WebRTC protocol stack, DTLS.
DTLS is used for all WebRTC data channels, and some media channels;
other media channels, however, used SDES for key exchange
and would go undetected by our script.
It also does not capture any features related to ICE/STUN/TURN.

We ran the script on a day's worth of network traffic
from Lawrence Berkeley National Laboratory.
The script found only seven DTLS handshakes,
with three unique client fingerprints and three unique server fingerprints.
This is less than we expected,
and indicates that there may not be all that much WebRTC
traffic in which to hide.
Part of the reason may be that Google Hangouts,
which we guessed would be the biggest contributor to WebRTC usage,
does not use DTLS.

\section{Future work}

We hope to expand this project by continuing to run the Bro script on other large traffic traces, improving the fingerprint found by this script, and creating automated scripts to fingerprint the STUN and TURN.
We ran the DTLS fingerprint script on only one day of traffic. We plan to run the script for longer periods of time, and on more traffic. The Bro script should also be improved. There is occasionally an anomaly present in DTLS connections,
which we haven't been able to explain yet,
where two consecutive client hellos are sent. These two packets are exactly the same, except for the sequence numbers, which are ``0'' then ``1''. Bro improperly handles these client hellos, leading to inaccurate results and missed logging of DTLS connections. Resolving this issue will go far in improving our knowledge of DTLS connections. Additionally, the STUN/TURN service should be analyzed on a large scale. This may involve creating another Bro script which checks for UDP connections to the STUN port (3478), then records the type of STUN packet and other features. 

Snowflake uses data channels,
while most of the applications we surveyed use media channels.
Though both types of channel are encrypted,
they are distinguishable because one uses DTLS while the other uses SRTP.
This leaves open the possibility that a censor could block
only data channels, without blocking WebRTC entirely,
resulting in smaller amount of false-positive-related cost to the censor.
It may be possible to abuse media channels to instead send binary data,
akin to what Houmansadr et~al. did with Freewave~\cite{Houmansadr2013a},
which modulated a data stream into an acoustic signal
to be transmitted over VoIP.
This would entail extra implementation difficulties
such as the need to layer a reliable transport layer
onto the unreliable media channels,
but it would make Snowflake's streams
not trivially distinguishable from those of other applications.

\section{Acknowledgments}

We wish to express thanks to Johanna Amann for help with Bro scripting,
to Vern Paxson for running our script against traffic,
and to Serene Han and Arlo Breault for comments.

\section{Availability}

Our DTLS fingerprinting Bro script is available from
\url{https://github.com/miagilepner/DTLS-fingerprint}.

{\footnotesize \bibliographystyle{acm}
\bibliography{dtls}}

\end{document}